\documentclass[epj, referee]{svjour} 
\usepackage{graphics}
\usepackage[english]{babel} 
\usepackage{epsfig,float,subfigure,amssymb,amsmath} 
\makeatletter

\newcommand{\beq}{\begin{equation}}
\newcommand{\eeq}{\end{equation}}
\newcommand{\bea}{\begin{eqnarray}}
\newcommand{\eea}{\end{eqnarray}}
\newcommand{\nonu}{\nonumber}

\DeclareMathOperator{\sech}{sech}

\begin{document}
\title{Kalb-Ramond excitations in a thick-brane scenario with dilaton}
\author{H. R. Christiansen\inst{1,2} \and M. S. Cunha\inst{1} }
\institute{Grupo de F\'{\i}sica Te\'orica, State University of Ceara (UECE), Av. Paranjana 1700, 60740-903 Fortaleza - CE, Brazil \and State University Vale do Acara\'{u}, Av. da Universidade 850, 62040-370 Sobral - CE, Brazil}
\date{Received: date / Revised version: date}
\abstract{
We compute the full spectrum and eigenstates of the
Kalb-Ramond field in a warped non-compact Randall-Sundrum -type
five-dimensional spacetime in which the ordinary four-dimensional braneworld
is represented by a sine-Gordon soliton.
This 3-brane solution is fully consistent with both the warped gravitational
field and bulk dilaton configurations. In such a  background we embed a bulk antisymmetric
tensor field and obtain, after reduction, an infinite tower of normalizable Kaluza-Klein
massive components along with a zero-mode.
The low lying mass eigenstates of the Kalb-Ramond field may be related to the axion pseudoscalar. This yields
phenomenological implications on the space of parameters, particularly on the dilaton coupling constant.
Both analytical and numerical results are given.
 }
\maketitle

\section{Introduction}\label{intro}

Field theoretic scenarios derived from string theory bring necessarily together the
standard model degrees of freedom and the gravitational field to interact
in some higher dimensional bulk space-time.

Extra dimensions have shown to be instrumental for solving fundamental
problems such as the hierarchy gap between the gravitational and the gauge coupling scales
leading to de Sitter (or de Sitter-like) geometries in four dimensions \cite{RS}.

Regarding standard gauge interactions, it is known that the electroweak  excitations are
deposited on D-branes by the
open strings ending on them \cite{polchinski}. Thus, normalized
{standard model gauge modes} are expected to be
localized on topological defects of lower dimensionality representing p-branes
in a low energy setup \cite{keha-tamva,cunhachris}. {On the same token, fermion fields
have also been found to be localized on the brane, as expected (see e.g. \cite{fermions}).}

Besides the standard model fields and particles
a series of entities not yet discovered
are the expected residuals left in quantum field theories emerging from a
low energy string inception. 
Among these the dilaton is a scalar field inevitably present along with the graviton.
Moreover, when gravitation is recovered from string theory, an antisymmetric  tensor appears 
together with graviton and dilaton in the Neveu-Schwarz bosonic sector of the low energy
string effective action \cite{ashoke}. Therefore, to start, we will be concerned with the low energy
interaction of these three fields plus a topological scalar representing our 3-brane world.

Motivations to consider the Kalb-Ramond field \cite{KR}
in a thick braneworld scenario are actually many.
It is known that since the graviton is a massless closed string excitation it can naturally propagate
at will across the whole space \cite{polchinski}. But then, one should also consider in the bulk other
fields related to closed strings. 
Two such  degrees of freedom are the above mentioned scalar dilaton and a
second-rank antisymmetric tensor usually met in supergravity theories \cite{superspace}.
In extended supergravity this tensor, associated with the Kalb-Ramond field,
becomes a factual component of the supergravity multiplet. 
Since massless modes are expected to be the most relevant in the ordinary world,
this sector of the low energy string theory is
likely responsible for most observable effects in cosmic phenomena \cite{GSW strings}.
{Although not a standard model component, the KR field has been argued to be localized on the
brane much in a similar way as the U(1) vector gauge field in \cite{CHRISTIANSEN,localized tensor}}.

In the last decade, the KR field has been related to torsion, a geometric property of spacetime
(such as curvature) that modifies the electromagnetic field  str\-ength through the affine connection
in the covariant derivative.
Using a generalized form of the Einstein-Cartan (EC) action, it has been shown \cite{indianos 99}
that the third-rank field-strength tensor of the KR field, $H_{\mu\nu\rho}$,  can be directly
identified with torsion to guarantee gauge symmetry. Interestingly, a parity violating term appears naturally\footnote{To cancel the $U(1)$ gauge anomaly and preserve $N = 1$ supersymmetry in the heterotic string theory the field strength \textit{H} is augmented by suitable Chern-Simons terms \textit{A F}.} if one considers a pseudotensorial extension of the affine connection.
Such a parity breaking gravitational interaction emerges not only from a theoretical viewpoint\footnote{
In the Einstein-Maxwell theory the electromagnetic field-strength reduces to the flat space expression for
the symmetric nature of the Christoffel connection \cite{puntingam}. When gravitation is described by the EC theory (a theory where the connection has an antisymmetric piece known as spacetime torsion) gauge symmetry is broken because torsion makes the electromagnetic field-strength (defined as the generally covariant curl of a 4-vector potential $A$)
no longer gauge invariant \cite{revmod 76}. Since it is a measurable quantity irrespective of the
background geometry the Maxwell field would not be allowed to minimally couple to EC theory.
However, when the spacetime torsion originates from a massless KR field with Chern-Simons terms added
parity is violated but gauge invariance preserved \cite{indianos 99}.} but also from observation.
For example, in \cite{indianos neutrino} it has been shown that a parity violating gravitation can flip the helicity of fermions providing a possible explanation to this neutrino problem. Likewise, the anisotropy in the cosmic microwave background radiation might be explained by a parity violating coupling of the electromagnetic field with a pseudoscalar \cite{CMB}, presumably the so-called axion.
Indeed the dual scalar of the pseudotensor component of the connection discussed above can
be identified with that particle.

It is worth mentioning that a completely antisymmetric torsion, as provided by the KR field,
can induce parity violation only in the matter (spin 1/2)
sector but not in the curvature or $U(1)$ gauge sector.
This is traced to the fact that such a tensor can always be expressed in terms of its dual  field, the axion
$\chi$, defined through the duality relation $H_{\mu\nu\rho}= \epsilon_{\mu\nu\rho\sigma} \partial^\sigma \chi$ (the solution for $\chi$ in flat space is linear in
the comoving time) \cite{indianos EPJ 2002}.
See \cite{axion KR} for early work relating the {axion} and the KR field (see also \cite{axion dilaton}).

In face of the above prospect,
in the present paper we will assume a massless five-dimensional KR field along with gravity.
However, since the breakdown of supersymmetry may result in the generation of mass for the
axion, we will also consider a Kaluza-Klein mechanism to endow the KR field with mass in ordinary space.
In this respect, it has been shown that dimensionally reduced $U(1)$ gauge fields are not
normalizable unless the effective coupling is modified by the dilaton \cite{youm1} so we shall
dynamically include the massless dilaton in the curved background defined for the KR field.
In other words, both brane and dilaton configurations will be geometrically consistent solutions of
a two scalar world action in a warped 5D spacetime. These configurations arise as topological kinks
of a sine-Gordon type potential function.

The paper is organized as follows. The framework is presented in the next section.
In Sect. \ref{sect action}
we study the five-dimensional equations of motion for the KR field and discuss a few
cases from a quantum-mechanical analog point of view.
Next, in Sect. \ref{sect heun} we cope with the general problem and compute spectra and
eigenfunctions for generic values of the parameters.
In Sect. \ref{remarks} we draw our conclusions.

\section{The model \label{sect model}}
We consider a 5D action where a bulk Kalb-Ramond field, $B_{NP}$,
is coupled to the dilaton, $\Pi$, in a warped space-time defined
by two metric functions ($\Lambda$ and $\Sigma$) and a brane field $\Phi$:
\bea
S = \int d^4x\,dy\,\sqrt{G}\ \left\{ 2M^{3}R_{(5)}-\frac{1}{2}(\partial\Phi)^{2}
-\frac{1}{2} (\partial\Pi)^{2}\right.\nonu\\\left.-\mathcal{V}(\Phi,\Pi)
+ \dfrac1{12}\, e^{-\lambda  \Pi }H_{MNP}H^{MNP} \right\}.
\label{action}
\eea
The tensor gauge field-strength is
$H_{MNP}=\partial_{M}B_{NP}+\partial_{P}B_{MN}+\partial_{N}B_{PM}$, $R_{(5)}$
is the Ricci scalar, and $M$ is the Planck mass in the bulk.
As usual, we employ Latin capitals in 5D and Greek lower case letters in the
four-dimensional slice.
We adopt the following ansatz for the metric
\begin{equation}
ds^{2}=G_{\mu\nu} dx^{\mu}dx^{\nu}+ e^{2\Sigma(y)} dy^{2},
\label{warpedmetric}
\end{equation}
where $G_{\mu\nu}=e^{2\Lambda(y)}\eta_{\mu\nu}$ and
diag$(\eta)=(-1, 1, 1, 1)$;  $G=-\det G_{MN}$. We will assume that
all the scalar fields depend just on the fifth coordinate, $y$ \cite{keha-tamva}.

Considering the scalars back reaction on the metric,
the equations of motion for $\Pi$, $\Phi$, $\Lambda$ and $\Sigma$ are
\bea &&\frac{1}{2}(\Phi^{\prime})^{2}+\frac{1}{2}(\Pi^{\prime})^{2}-
e^{2\Sigma(y)}\mathcal{V}(\Phi,\Pi)=24M^{3}(\Lambda^{\prime})^{2},\nonumber\\
&&\frac{1}{2}(\Phi^{\prime})^{2}+\frac{1}{2}(\Pi^{\prime})^{2}+e^{2\Sigma(y)}\mathcal{V}(\Phi,\Pi)=
-12M^{3}\Lambda^{\prime\prime}\nonumber\\
&&-24M^{3}(\Lambda^{\prime})^{2}+12M^{3}\Lambda^{\prime}\Sigma^{\prime},\label{motion1}
\eea and  \bea
\Phi^{\prime\prime}+(4\Lambda^{\prime}-\Sigma^{\prime})\Phi^{\prime}=
e^{2\Sigma}\ \frac{\partial\mathcal{V}}
{\partial\Phi},\nonumber\\
\Pi^{\prime\prime}+(4\Lambda^{\prime}-\Sigma^{\prime})\Pi^{\prime}=
e^{2\Sigma}\ \frac{\partial\mathcal{V}}{\partial\Pi}.
\label{motion2} \eea
where the prime means derivative with respect to $y$.

The potential functional $\mathcal{{V}}(\Phi, \Pi)$ stems from a supergravity
motivated definition \cite{supergravity} which can be applied to non-supersymmetric systems.
Based on a standard sine-Gordon potential, the functional $V(\Phi)=b^{-2}\ (1-\cos (b\,\Phi))$
gets deformed in order to include the effects of the warped metric and the dilaton.
The final expression
\bea
\mathcal{{V}}(\Phi, \Pi)=e^{({\Pi}/{\sqrt{12M^{3}}})}& &
\left(\frac 2{b^2}\sin^2(\frac b 2\,\Phi+\frac{\pi}2)-\right.\nonu\\
& & \left.\frac{5}{2M^3 b^4}\cos^{2}(\frac b 2\,\Phi+\frac{\pi}2)\right),
\label{newpotential}
\eea
is found after some algebra
by looking for the consistency of the equations of motion (\ref{motion1}) and (\ref{motion2}) \cite{CHRISTIANSEN} and symmetrizing the bounce eq.(\ref{bounce}) with respect to $y=0$.
The solutions to the equations for $\Phi$ and $\Pi$ represent
the braneworld and the dilaton final configurations respectively. They are
consistent altogether with the metric functions $\Lambda$ and $\Sigma$.
Indeed, the following solution, interpolating vacua and kinking on our 4D-world slice
\beq
\Phi=\frac 4 b \ (\arctan\ e^y - \frac{\pi}4)\label{bounce}\eeq
is compatible with a dilaton configuration
\beq
\Pi= \frac 1 {\sqrt{3M^3} b^2} \ln\cosh y\label{dilaton},\eeq
in a gravitational field given by
\beq
\Lambda=-\frac 1 {{3M^3} b^2} \ln\cosh y,
\label{warp}\eeq
and
\beq \Sigma=-\frac 1 {{12M^3} b^2} \ln\cosh y. \label{sigma}\eeq
With this set of solutions the scalar action is finite
provided $\lambda$ is above a critical value $\lambda_0 = -{17}/{4\sqrt{3M^3}}$ \cite{CHRISTIANSEN}.
By means of the warping functions (\ref{warp}) and (\ref{sigma}), the effective
four-dimensional Planck scale can be defined by
\bea M_P^2 &=& M^3\int_{-\infty}^\infty dy \ e^{4\Lambda(y)+\Sigma(y)}\nonumber\\
& =&M^3\int_{-\infty}^\infty dy \ (\sech y)^{{C}/{M^3}}
\eea
where $C=17/12b^2$. After integration, we obtain a (finite) close expression for the effective Planck scale
as a function of the 5D Planck scale
\bea M_P^2=M^3\, \frac{\sqrt{\pi}\,\,\Gamma\!\left( \frac{C}{2M^3} \right)}{\Gamma\!\left(\frac12+ \frac{C}{2M^3}\right)}.\label{Mb}\eea

By studying the fluctuations of the metric about the background configuration, eqs.(\ref{bounce})-(\ref{sigma}),
it is possible to see that this model supports a normalizable massless graviton (zero-mode of the gravitational
field) localized on the membrane  \cite{keha-tamva}. 

\section{Field equations of the KR field \label{sect action}}

Since the axion field is supposed to be extremely diluted in space
it is likely that the Kalb-Ramond field would not contribute significantly
to the geometrical background.
Indeed, in a Randall-Sundrum scenario it has been
explicitly shown that the Kalb-Ramond energy density has
only a tiny value $\sim 10^{-62}$ \cite{KR RS}.
Thus, we can safely study the behavior of propagating modes in the
topological background configuration above found.

The equations of motion for $B_{MN}$ in the bulk are given by
\beq
\frac 1{\sqrt{G}}\,\partial_M (G^{MR}G^{NS}G^{PQ} H_{RSQ}\sqrt{G}
e^{-\lambda  \Pi(y)})=0. \label{more motion}
\eeq
%
In order to solve these equations, we adopt the following gauge choices
\beq B^{\mu 5}=0,\
\partial_{\mu}B^{\mu\nu}=0.\label{gauge choice}\eeq
Next, we perform the usual Kaluza-Klein decomposition as follows
\beq B^{\mu\nu}(x, y)=\sum_n\,b_n^{\mu\nu}(x)w_n(y)\label{decomposition}\eeq
yielding
\beq
\left[\Box +\left(\frac 1{w_n\ g} \partial_5(g\partial^5 w_n)\right)\right]\ b_n^{\mu\nu}=0
\eeq
with $g(y)=e^{4\Lambda+\Sigma}\exp[-\lambda\Pi]$.
The Kaluza-Klein spectrum of the Kalb-Ramond field is then given by
\beq
[g^{-1}\partial_5 g\partial^5  + m_n^2]\ w_n=0,
\eeq
where $m_n^2$ is the 4D squared boson mass
of the tensor field $n$th-mode, satisfying $p_n^2=-m_n^2$.
Operating further, the $y$-dependent equation reads
\bea
{6M^3b^2}w_n''(y)-{(15/2+2\lambda\sqrt{3M^3})}\tanh y\ w_n'(y) +\nonumber\\
 m_n^2{6M^3b^2}\sech^{1/6M^3b^2}y\ w_n(y)=0,\label{massive modes 1}\eea
where the prime means $\partial/\partial y$.

{In contrast to Ref. \cite{CHRISTIANSEN}, here we adopt a
different relation between the fundamental parameters $M$ and $b$
in order to analyze a new equation of motion for the gauge field.
Now, the last term of eq.(\ref{massive modes 1})
involves the function $\sech^6y$ instead of $\sech^2y$
that was considered in \cite{CHRISTIANSEN}.
This, together with a different constant factor in the two last terms,
imposes the necessity of solving a new differential
equation which brings about different spectra and the eigenfunctions
associated.}

With the choice $36M^3\,b^2=1$
eq.(\ref{massive modes 1}) can be written as
\beq
w_n''(y)+\tilde{c} \tanh y\ w_n'(y)
  + m_n^2\sech^{6}y\ w_n(y)=0, \label{massive modes}
\eeq
where $\tilde{c} =-3(15+4\lambda\sqrt{3M^3})$.  For simplicity, we
drop the subindexes of the propagation modes in what follows. We will solve
this equation exactly in what follows.
\ {}
\ {}
\subsection{Special cases \label{sub quantum analog}}

By means of the transformation
\beq  w(y)=e^{-\alpha \Lambda}\,W(z),\ \ \ \frac{dz}{dy}= e^{\Lambda/4}\label{change variables} \eeq
we can turn eq.(\ref{massive modes 1}) into
\beq \left[-\frac{d^2}{dz^2}+ \mathfrak{V}(z)\right]W(z)=m^2W(z),\label{schrodinger}\eeq
where $\mathfrak{V}(z)={\alpha}\,e^{-\Lambda/2}(\Lambda''+({\alpha}-\frac 1 4)\ \Lambda'{~}^2)$
and $\alpha=2+\lambda\sqrt{3M^3}/2$, which is
a Schrodinger-like equation in the variable $z\in(-1,1)$\ (see e.g. \cite{RS,keha-tamva}).

\begin{figure}[hb]
\centering
\includegraphics[width=7.5cm,height=5.5cm]{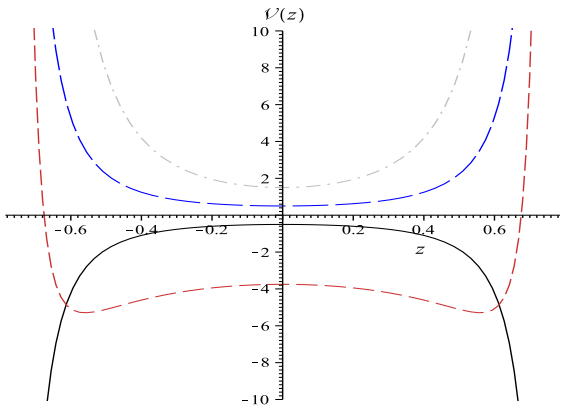}
\caption{\label{pot_z_a6} Potential $\mathfrak{V}(z)$ numerically computed for some values of the parameters: $\gamma=1/2$ (solid black line), $\gamma=30/8$
(dashed red line), $\gamma=-1/2$ (long-dashed blue line), {and $\gamma_0=-3/2$ (dot-dashed grey line),
corresponding to critical  $\lambda_0$}. Note the divergency at $z \rightarrow \pm \pi/4$.}
\end{figure}

In the present case we obtain
\beq
z(y) = \arctan (\tanh\frac{y}{2}) + \frac{1}{2} \sech \!y \tanh \!y ,\label{z(y)}
\eeq
\beq
w(y)=\cosh^{\gamma}\!y \,W(z), \label{uy}
\eeq
and
\beq %
\mathfrak{V}(z(y)) = -\gamma \cosh^4 \!y  \left [1+(3-\gamma) \sinh^2 \!y \right ], \label{potencial_a6}
\eeq %
where $\gamma=12\,\alpha$ and $\tilde{c}=3-2\gamma$. {Both the change of variables (\ref{z(y)}), (\ref{uy}) and the potential function (\ref{potencial_a6}) are completely different
from those in \cite{CHRISTIANSEN}.} In the present case the solution is more difficult, but
although we cannot analytically invert eq.(\ref{z(y)}), we can
do it numerically and compute $\mathfrak{V}(y(z))$ for several values of $\gamma$,
see Fig.\ref{pot_z_a6}.

%
%

\subsubsection{The case $\gamma$=0 ($\tilde{c}= 3$) \label{subsub tilde3}}

\begin{figure}[ht]
\centering
\includegraphics[width=7.5cm,height=5.5cm]{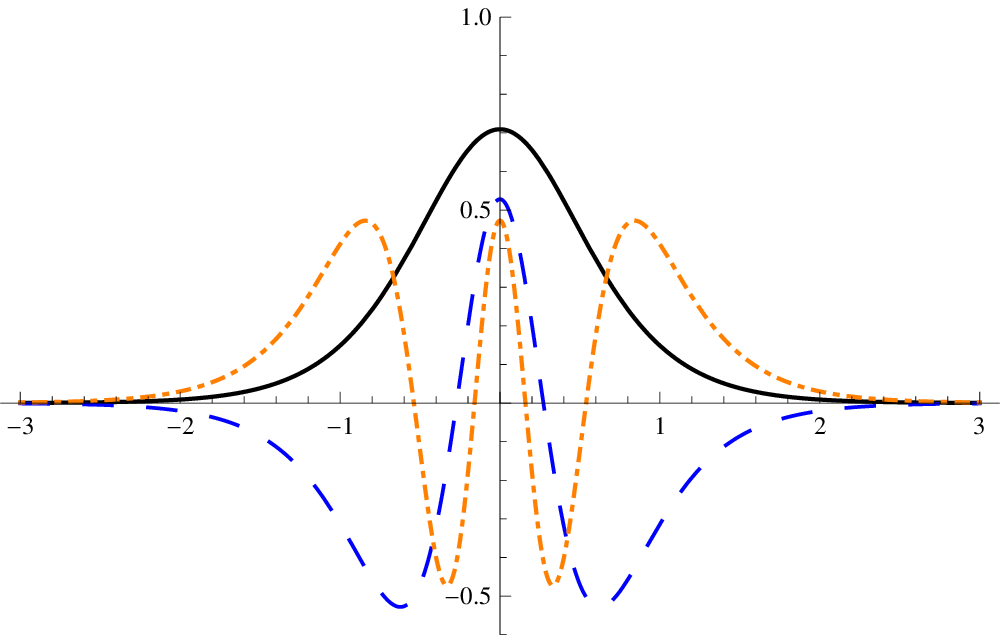}
\caption{\label{a6_c3_par_y_sym} First normalized symmetric  solutions,
Eq. (\ref{wy1_c1_4}), for $\tilde{c} =3$:
$m= 2$ (solid  black line), $m=6$ (dashed blue line), $m=10$ (dot-dashed orange line). (Color figure online)}
\includegraphics[width=7.5cm,height=5.5cm]{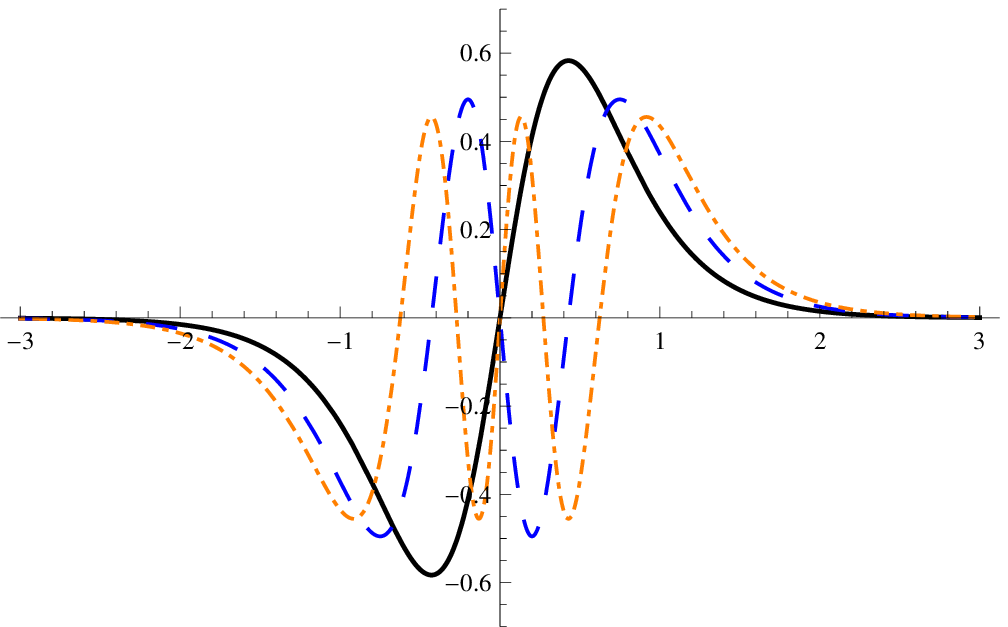}
\caption{\label{a6_c3_impar_y_asym} First normalized antisymmetric solutions,
Eq. (\ref{wy2_c1_4}), for $\tilde{c} =3$:
$m=4$ (solid black line), $m=8$ (dashed  blue line), $m=12$ (dot-dashed orange line). (Color figure online)}
\end{figure}

In this case, the potential function (\ref{potencial_a6}) seems trivial, but
in fact, for $z=\pm \pi/4$ we have $y \rightarrow \pm \infty$
and therefore solutions to eq. (\ref{schrodinger}) must be null in $|z|\geq \pi/4$.
The potential in $z$ thus corresponds to an infinite square-well of width $\pi/2$.
The normalized solutions of the Schrodinger equation (\ref{schrodinger})
can be analytically obtained
\beq
W(z) = \frac{2}{\sqrt{\pi}} \cos((4n+2)\ z) + \frac{2}{\sqrt{\pi}} \sin(4n\ z). \label{Wz_c1_4}
\eeq
Here, the mass spectrum is given by $m=2(n+1)$ where the parity of $n$ is the parity of the associated solution.
In the actual $y$ space these solutions read
\bea
w^{(1)}(y) = \frac{2b_1}{\sqrt{\pi}} \cos((4n+2)( \arctan (\tanh\,\frac{y}{2})~~~~~~~~~~ \nonumber\\
+ \frac{1}{2} \sech \,y \tanh \,y))~~~\label{wy1_c1_4}
\eea
and
\bea
w^{(2)}(y) = \frac{2b_2}{\sqrt{\pi}} \sin(4n \arctan (\tanh\,\frac{y}{2}) ~~~~~~~~~~~~~~~~~ \nonumber\\
+ \frac{1}{2} \sech \,y \tanh \,y)).~  \label{wy2_c1_4}
\eea
where $b_1, b_2$ are normalization constants.
Solutions (\ref{wy1_c1_4}) and (\ref{wy2_c1_4}) for the first values of the spectrum
are shown in Figs. \ref{a6_c3_par_y_sym} and \ref{a6_c3_impar_y_asym} respectively.
%
%
%

\subsubsection{The $\gamma$=1/2, ($\tilde{c}=2$) case \label{subsub tilde2}}
%
%
\begin{figure}[ht]
\centering
\includegraphics[width=7.5cm,height=5.5cm]{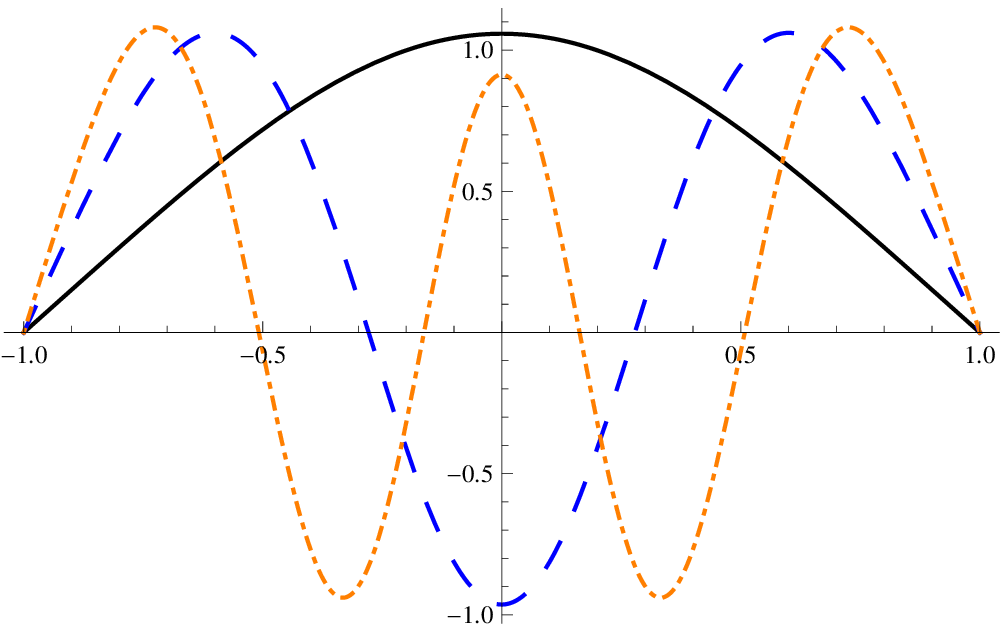}
\caption{\label{a6_c2_par} Symmetric solutions Eq. (\ref{parabolicas}) for $\tilde{c} =2$:
 $m=1.68159532$ (solid  black line), $m=5.66985735$ (dashed blue line), $m=9.66824247$ (dot-dashed orange line).
 (Color figure online)}
\includegraphics[width=7.5cm,height=5.5cm]{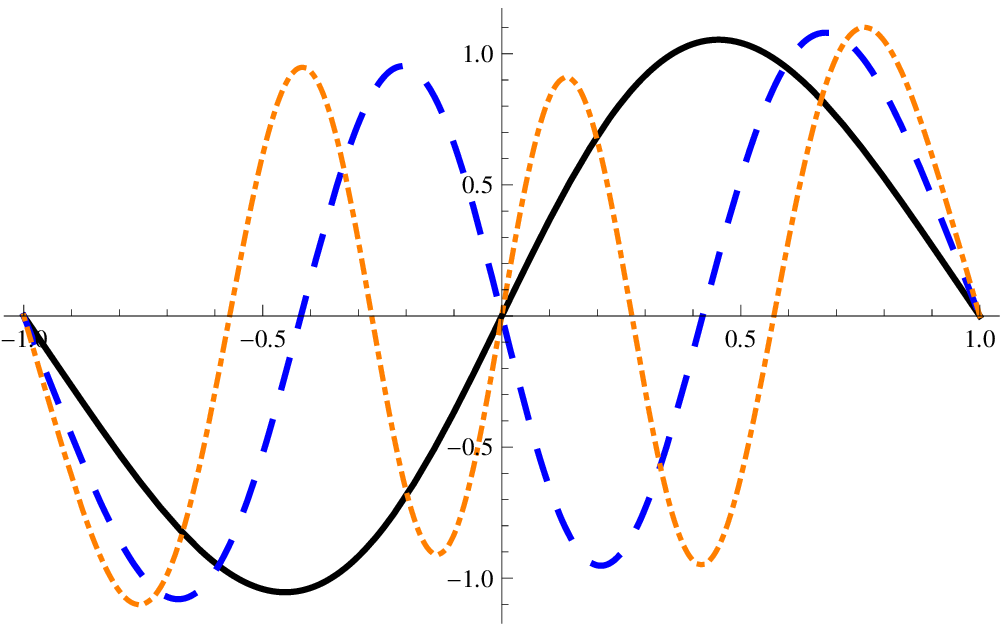}
\caption{\label{a6_c2_impar} Antisymmetric solutions Eq. (\ref{parabolicas}) for $\tilde{c} =2$:
$m=3.67229037$ (solid black line), $m=7.668808762$ (dashed  blue line), $m=11.667894313$ (dot-dashed orange line). (Color figure online)}
\end{figure}
%
By means of the following transformation
\beq  {x}= \sqrt{m}\tanh(y)
\eeq
we can rewrite  eq. (\ref{massive modes}) as
\beq
w''(x) + (\tilde{c}-2) \frac{x}{m-x^2} w'(x) + (m-x^2) w(x) =0. \label{eqx}
\eeq%
%
%
Now, with  $\tilde{c}=2$ we get
\beq -w''( {x}) +  {x}^2 w( {x}) = m\ w( {x})
\label{tildex} \eeq
This is the equation of motion of a nonrelativistic quantum particle of energy $m$
in a harmonic potential $V({x}) = {x}^2$ for $-\sqrt{m}<{x}<\sqrt{m}\ $
($V({x}) = \infty$ for $|{x}| \geq \sqrt{m}$ and thus $w({x}) =
0$ in this region).

With $x\rightarrow \frac{x}{\sqrt2}$, it reads as a standard
parabolic cylinder differential equation
\beq w''( {x}) +  (-\frac{x^2}4+\frac m2)\, w({x}) = 0
\label{paracyl} \eeq
with analytical even and odd solutions given by
\bea
w_1(x)=e^{-x^2/4}\ _1F_1(-\frac m4+\frac 14;\frac 12;\frac{x^2}2),\nonu\\
w_2(x)=xe^{-x^2/4}\ _1F_1(-\frac m4+\frac 34;\frac 32;\frac{x^2}2),
\eea
\cite{ABRAMOWITZ} where  $_1F_1$ are confluent hypergeometric functions.

The solution can also be written in terms of parabolic cylinder
functions
\bea && D_\nu(x) = \frac{2^{\nu/2} \sqrt{\pi}}{\Gamma(\frac{1-\nu}{2})} %
\left[1- \frac{1}{2}\,(1 + 2\nu)\, \frac{x^2}{2!}
 + \frac{1}{4}\,(3 + 4\nu + 4\nu^2)\,\frac{x^4}{4!}\right. \nonumber\\
&& \left.  - \frac{1}{8}\,(15 + 34\nu + 12\nu^2 + 8\nu^3)\,\frac{x^6}{6!} +
 \mathcal{O}(x^{8}) \right]\nonumber \\ 
&&- \frac{2^{\frac{\nu + 1}{2}} \sqrt{\pi}}{\Gamma(-\frac{\nu}{2})} %
\left[x - \frac{1}{2}\,(1 + 2 \nu)\,\frac{x^3}{3!} %
+\frac{1}{4}\,(7 + 4 \nu + 4 \nu^2)\,\frac{x^5}{5!}\right. \nonumber \\
&& -\left.\frac{1}{8}\,(27 + 58 \nu + 12 \nu^2 + 8 \nu^3)\,\frac{x^7}{7!}+
\mathcal{O}(x^9)\right]
\eea
as
\beq
w(\tilde{x}) = A_1\ D_{\frac{m-1}{2}}(\sqrt{2m}~\tilde{x})+A_2\ D_{\frac{m-1}{2}}(-\sqrt{2m} \,\tilde{x}), \label{parabolicas}
\eeq %
where $A_1 = A_2 = A/\sqrt{2}$ for symmetric solutions and $A_1 = -A_2 = A$ for the
anti-symmetric ones. Here, $\tilde{x}=x/\sqrt{2m}$.
For simplicity, the value of the normalization constant can be approximated by
$A\simeq 1/\sqrt{((m-m_0)/2)!}$, where $m_0=1.556$ for even solutions
and $m_0=1.581$ for odd solutions.
The mass eigenvalues can also be well approximated by  $m\simeq 2n+5/3$, with $n \in \mathbb{N}$, see Figs. \ref{a6_c2_par} and \ref{a6_c2_impar}.

\begin{figure}[ht]
\centering
\includegraphics[width=7.5cm,height=5.5cm]{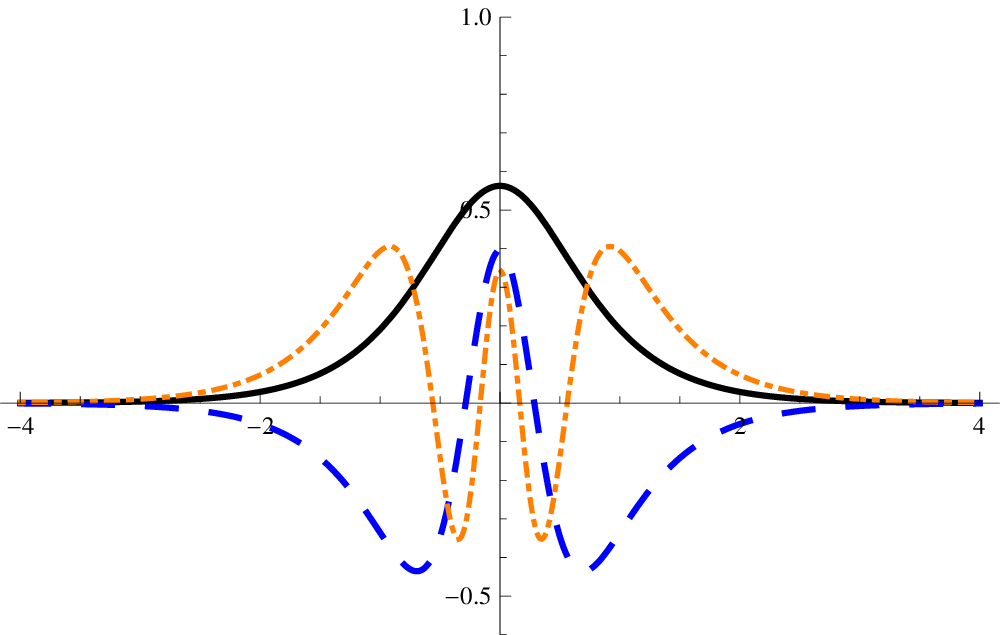}
\caption{\label{a6_c2_par_y_sym} Symmetric solutions $w^{(1)}(y)$, Eq. (\ref{parabolicas_y1}),
for $\tilde{c} =2$:
$m=1.68159532$ (solid  black line), $m=5.66985735$ (dashed blue line), $m=9.66824247$ (dot-dashed orange line).
 (Color figure online)}
\includegraphics[width=7.5cm,height=5.5cm]{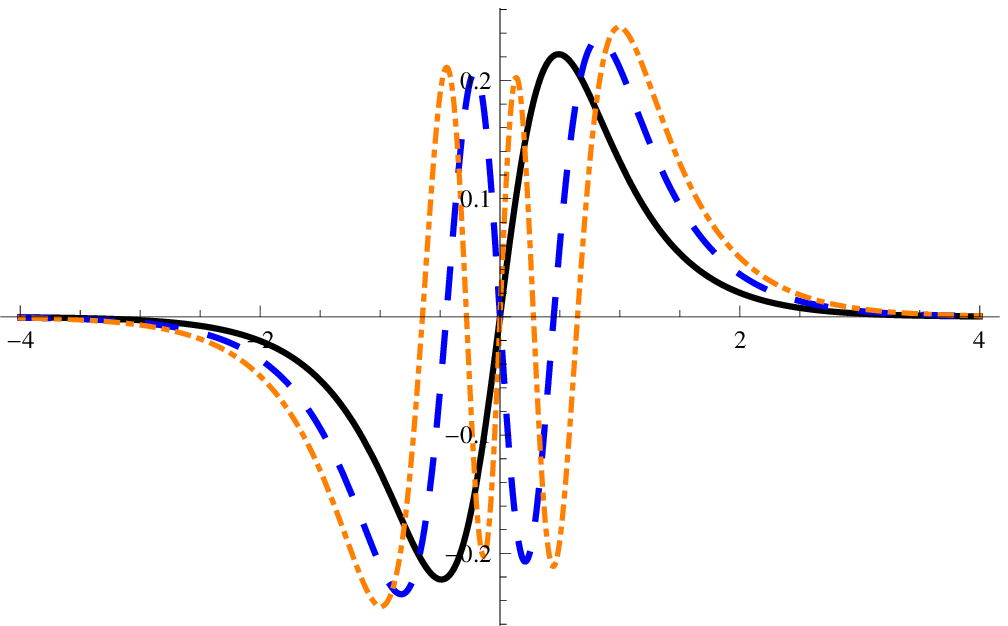}
\caption{\label{a6_c2_impar_y_asym} Antisymmetric solutions $w^{(2)}(y)$, Eq. (\ref{parabolicas_y2}),
for $\tilde{c}=2$:
$m=3.67229037$ (solid black line), $m=7.668808762$ (dashed  blue line), $m=11.667894313$ (dot-dashed orange line). (Color figure online)}
\end{figure}

In terms of $y$,  solutions (\ref{parabolicas}) read
\bea %
w^{(1)}(y) = c_1\frac{A}{\sqrt{2}}\left[ D_{\frac{m-1}{2}}(\sqrt{2m}~\tanh(y)) ~~~~~~~~~~~~~~~~~~~~~\right. \nonumber\\
\left. + D_{\frac{m-1}{2}}(-\sqrt{2m} \,\tanh(y)) \right]~~~~ %
\label{parabolicas_y1}
\eea %
\bea %
w^{(2)}(y) = c_2A \left[ D_{\frac{m-1}{2}}(\sqrt{2m}~\tanh(y))~~~~~~~~~~~~~~~~~\right. \nonumber \\
\left. - D_{\frac{m-1}{2}}(-\sqrt{2m} \,\tanh(y)) \right].
\label{parabolicas_y2}
\eea %
where $c_1, c_2$ are new normalization constants. See Figs.  \ref{a6_c2_par_y_sym} and \ref{a6_c2_impar_y_asym}.
%

Looking at eq.(\ref{potencial_a6}),  cases  like $\gamma=3$ and $\gamma=2$ result
in simple forms of $\mathfrak{V}(z(y))$ seemingly affordable. However
these do not bring about exact solutions in the Schrodinger
approach because such expressions cannot be analytically inverted in terms of $z$ (see eq.\ref{z(y)}).


\section{The general solution \label{sect heun}}

In order to obtain the general solution to Eq. (\ref{massive modes}) we perform
the following transformation
\beq z = \tanh^2 \!y . \eeq
Now,  Eq. (\ref{massive modes}) becomes
%
\bea
w''(z) + \left( \frac{1/2}{z} + \frac{1-\tilde{c}/2}{z-1} \right) w'(z)
~~~~~~~~~~~~~~~~~~~~~~~~~~~~~\nonumber \\
 + \frac{m^2}{4} \left(\frac{1-z}{z}\right)w(z) = 0 ~~~~~ \label{eqUa4b}
\eea
which has two Fuchsian points,  at $z=0$ and $z=1$,
and an irregular singularity, at $z=\infty$. If we now define
\beq
w(z)=e^{\mu z} W(z)
\eeq
we get
\bea
W''(z) + \left(2\mu +\frac{1/2}{z}+\frac{1-c/2}{z-1} \right) W'(z) + \frac{1}{z(z-1)}
\times ~~~~~ &&  \nonumber\\
\left[ z^2 \left(\mu^2 -\frac{m^2}{4} \right) +
z\left(-\mu^2 + \frac{\mu}{2}+\mu (1-\tilde{c}/2)+\frac{m^2}{2} \right) \right. ~~ \nonumber \\
\left. - \frac{\mu}{2} -\frac{m^2}{4} \right] W(z) = 0.~~~~~~~&&
\eea
By choosing $\mu = m/2$ we thus obtain
\bea
W''(z) &+& \left(m +\frac{1/2}{z}+ \frac{1-\tilde{c}/2}{z-1} \right) W'(z)
~~~~~~~~~~~~ ~~~~~~~~~\nonumber \\
&+& \frac{m}{4} \,\frac{(m+1)(z-1)+2-\tilde{c}}{z(z-1)} \,\, W(z) =0. \label{eqUz}
\eea
We can verify that this is a particular case of the canonical
non-symmetrical confluent Heun equation \cite{HEUN,heunDE}
as given in \cite{HOUNKONNOU,heunDE,CLEB-TERRAB}
\bea
Hc''(z) + \left(\alpha + \frac{\beta + 1}{z} + \frac{\gamma + 1}{z-1} \right)Hc'(z) + ~~~~~~~~~~  \nonumber \\
+ \frac{1}{z(z-1)}\left[ [\delta + \frac{\alpha}{2}(\beta + \gamma +2)]z +\eta ~~~~~~~~\right. \nonumber \\
\left. + \frac{\beta}{2}  + \frac{1}{2} (\gamma - \alpha)(\beta+1) \right] Hc(z) = 0,~~~~~~~~~ \label{heunc}
\eea
whose solutions around $z=0$ are given by
\bea
H^{(1)}&=& Hc\,(\alpha,\beta,\gamma,\delta,\eta; z) \\
H^{(2)}&=& z^{-\beta}\,Hc\,(\alpha,-\beta ,\gamma , \delta, \eta; z).%
\eea
%
%
\begin{figure}[ht]
\centering
\includegraphics[width=7.5cm,height=5.5cm]{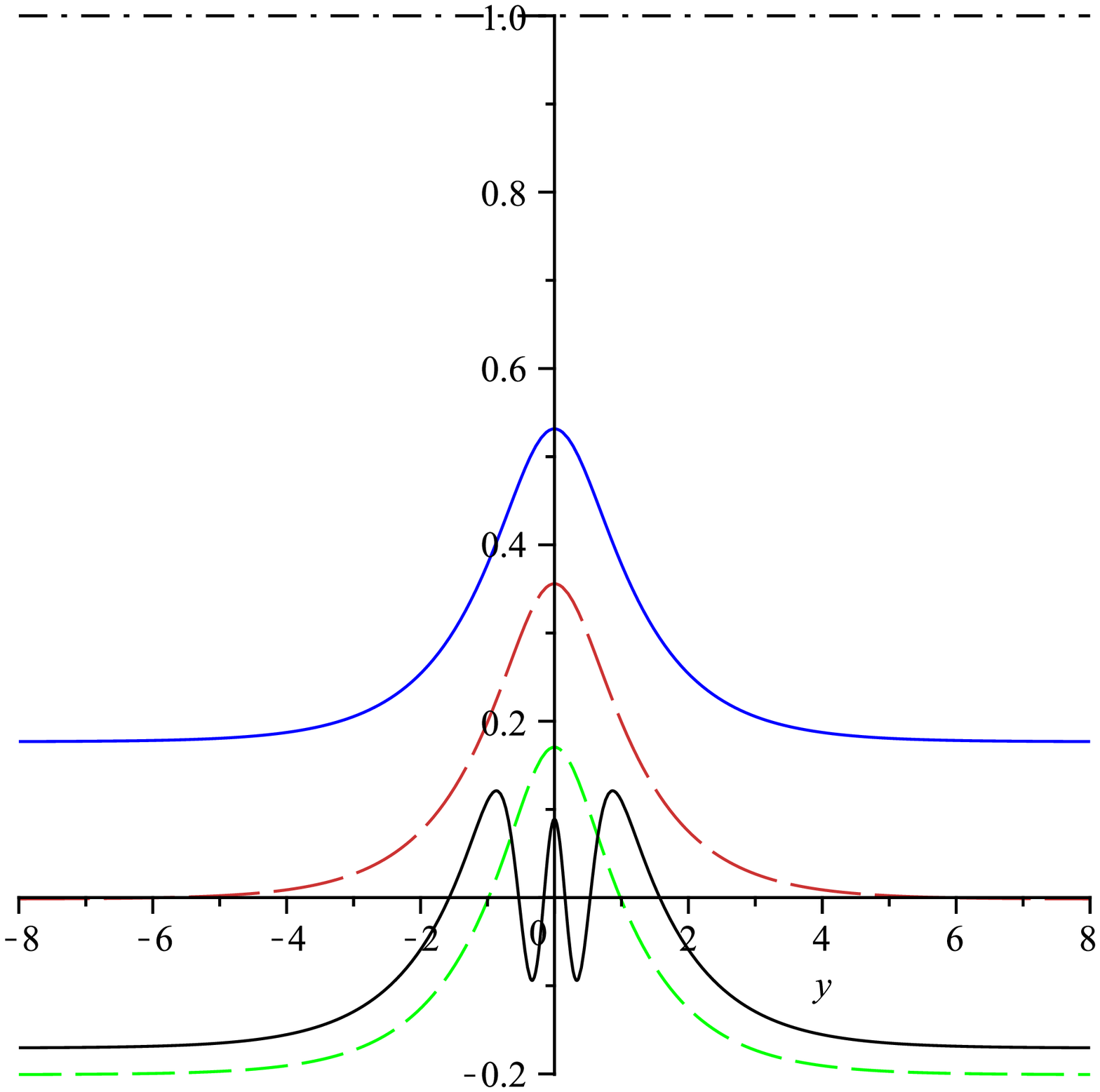}
\caption{\label{a_6_1 sym} Symmetric solutions normalized, Eq. (\ref{hc_a4_s}), for $\tilde{c} = 1$;
 $m=0$ (dash-dotted black line), $m=1$ (solid  blue line), $m=1.25$ (long-dashed red line),
 $m=2$ (dashed green line), and $m=10$ (solid black line).}
\includegraphics[width=7.5cm,height=5.5cm]{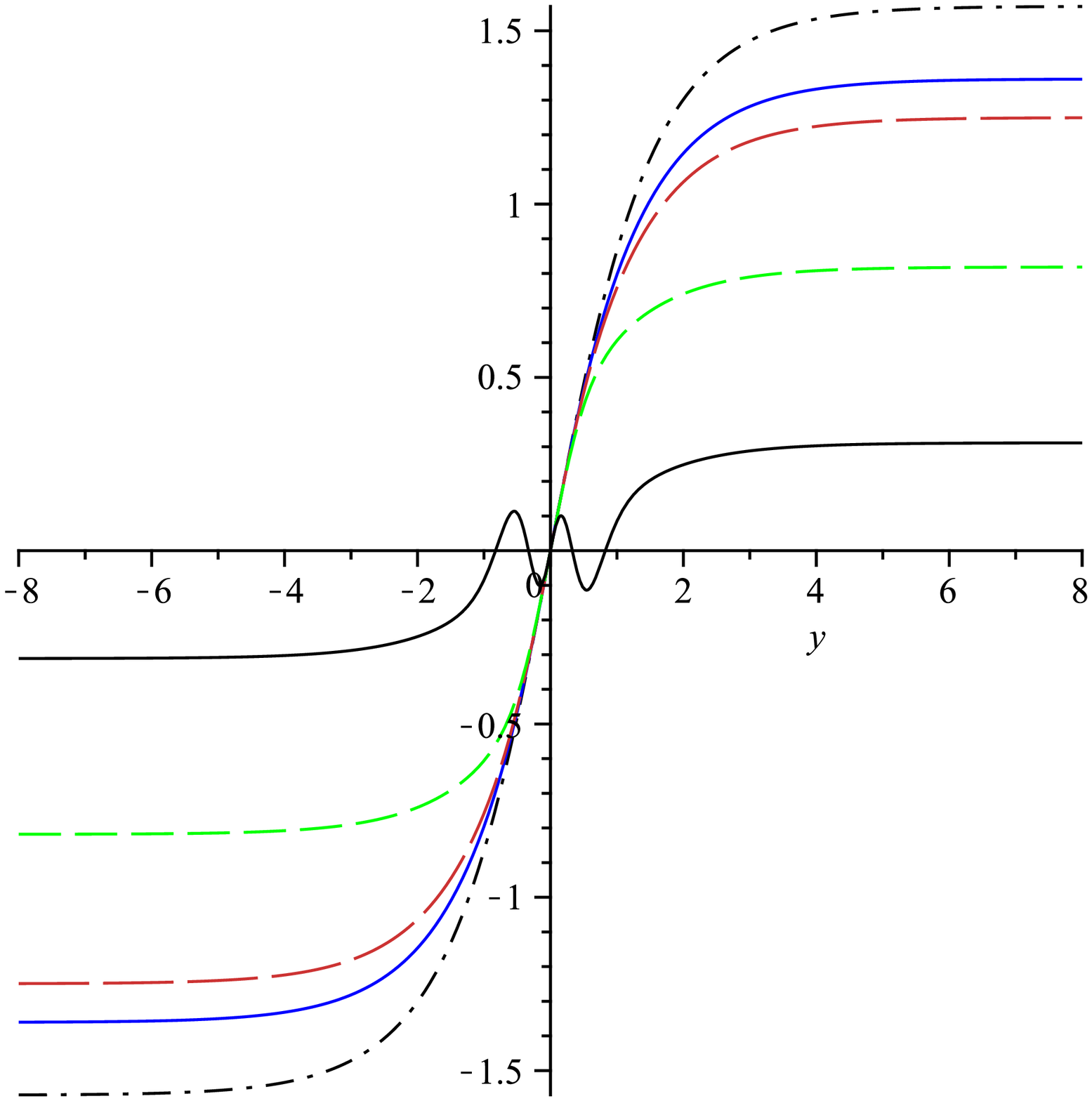}
\caption{\label{a_6_1 asym} Asymmetric solutions, Eq. (\ref{hc_a4_a}), for $\tilde{c} = 1$;
 $m=0$ (dash-dotted black line), $m=1$ (solid  blue line), $m=2$ (long-dashed red line), $m=5$ (dashed green line), and $m=10$ (black line).}
\end{figure}
In the region of interest, $z<1$,  regular local solutions around $z=0$ are defined by
the Heun series
\beq
Hc(z) = \sum_{n=0}^\infty d_n \, z^n,
\eeq
where the constants $d_n$ (with $d_{-1}=0$ and $d_0=1$)
are determined by the three-term recurrence relation \cite{FIZIEV}
\bea
A_n d_n= B_n d_{n-1}+C_n d_{n-2} \label{recurrence}
\eea
with
\bea
A_n &=& 1+\frac{\beta}{n} \rightarrow 1- \frac{1}{2n} \\
B_n &=& 1+ \frac{-\alpha + \beta + \gamma-1}{n} \nonumber \\ &+&\frac{\eta+(\alpha-\beta-\gamma)/2-\alpha \beta/2+\beta \gamma/2}{n^2} \nonumber \\
&\rightarrow & 1 + \frac{m-\tilde{c}/2 -3/2}{n}\nonumber\\
&+&\frac{\tilde{c}/2+1/2-m^2/4+3m/4}{n^2} \\
C_n &=& \frac{1}{n^2} \left(\delta+\frac{\alpha(\beta+\gamma)}{2}+\alpha (n-1) \right) \nonumber \\ &\rightarrow & \frac{m^2}{4n^2}+\frac{m}{n^2}\left(n-\tilde{c}- \frac{5}{4} \right).
\eea
We can identify $\alpha=m$, $\beta=-1/2$, $\gamma =-\tilde{c}/2$,
$\delta=m^2/4$, and $\eta=\tilde{c}/8 +1/4-m^2/4$
by comparing Eqs. (\ref{eqUz}) and (\ref{heunc}).
The solutions to Eq. (\ref{massive modes}) are therefore given by
\bea
w^{(1)}(y) = e^{\frac{m}{2}\tanh^2\!y} \times ~~~~~~~~~~~~~~~~~~~~~~~~~~~~~~~~~~~~~~~~~~~~~~~ \nonumber \\
Hc \left( m,-\frac{1}{2},-\frac{\tilde{c}}{2},\frac{m^2}{4},\frac{1}{4}+\frac{\tilde{c}}{8}-\frac{m^2}{4}; \, \tanh^2 \!y \right)~~~~~~ \label{hc_a4_s}\\
w^{(2)}(y) = e^{\frac{m}{2}\tanh^2\!y}  \tanh\!y \times ~~~~~~~~~~~~~~~~~~~~~~~~~~~~~~~~~~~~~~\nonumber \\
 Hc \left( m,\frac{1}{2},-\frac{\tilde{c}}{2},\frac{m^2}{4},\frac{1}{4}+\frac{\tilde{c}}{8}-\frac{m^2}{4}; \, \tanh^2 \!y \right)~~~~~~~~ \label{hc_a4_a}
\eea
for arbitrary values of $\tilde{c}$, i.e. for any value of the dilaton coupling constant.

\begin{figure}[ht]
\centering
\includegraphics[width=7.5cm,height=5.5cm]{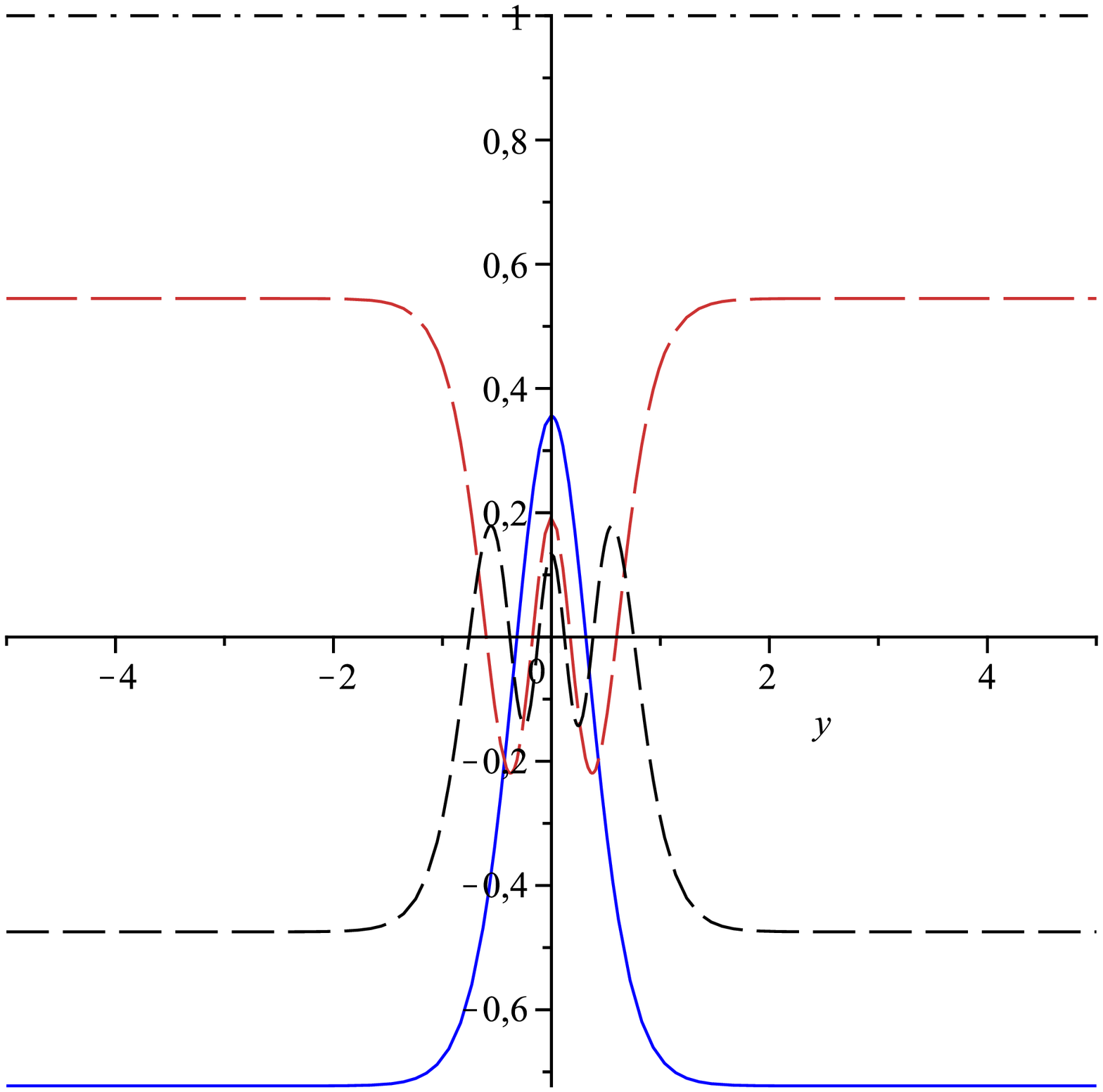}
\caption{\label{a_6_sym} Symmetric  solutions, Eq. (\ref{hc_a4_s}), for $\tilde{c} = -1$ ($\gamma=2$);
 $m=0$ (dash-dotted black line), $m=5.01256460$ (solid  blue line), $m=9.11222614$ (long-dashed red line), and $m=13.15839916$ (dashed black line).}
\includegraphics[width=7.5cm,height=5.5cm]{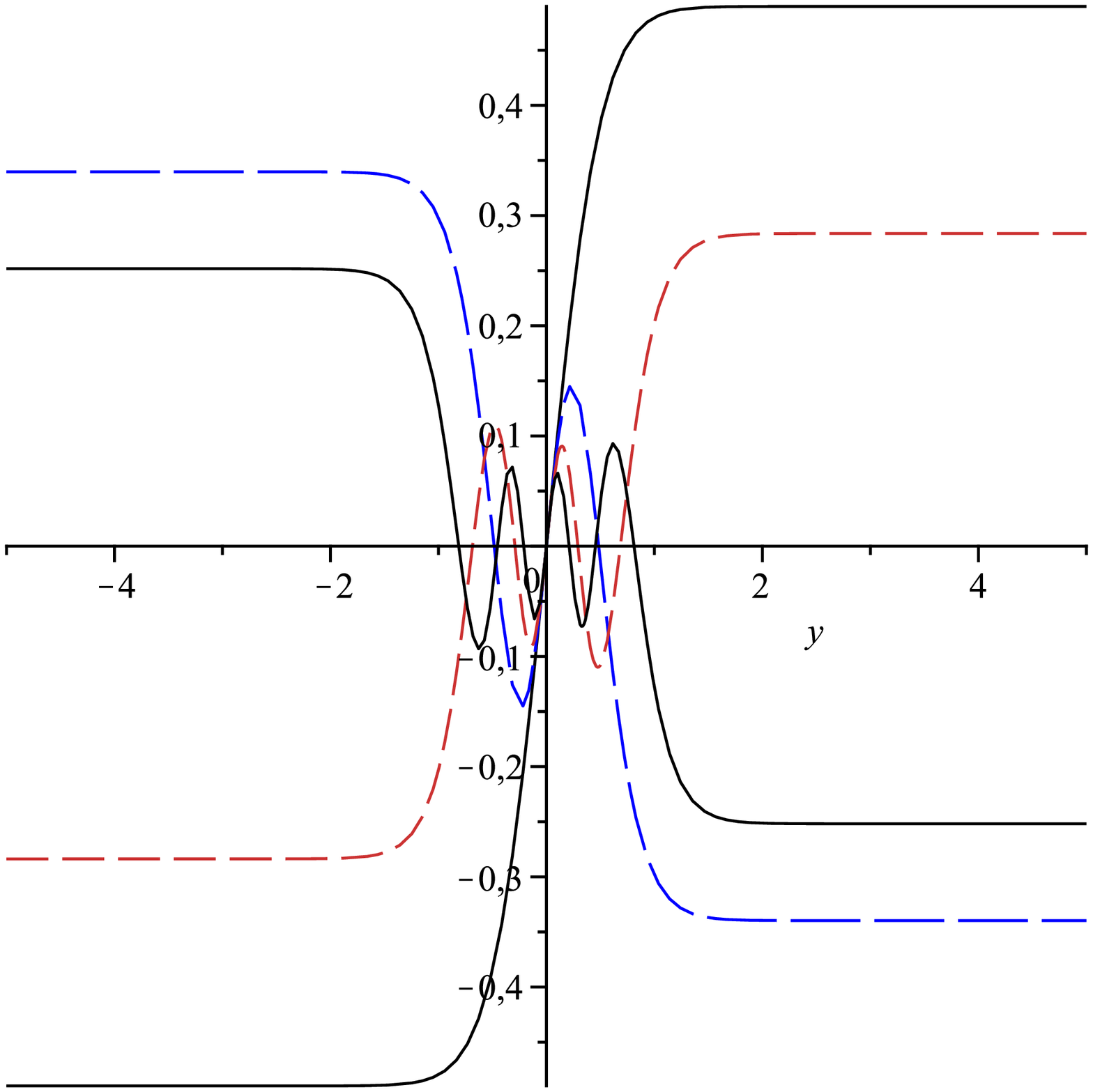}
\caption{\label{a_6_asym} Asymmetric solutions, Eq. (\ref{hc_a4_a}), for $\tilde{c} = -1$ ($\gamma=2$);
 $m=2.88880921142$ (dash-dotted black line), $m=7.074052696$ (solid  blue line), $m=11.138717717$ (long-dashed red line), and $m=15.173713228$ (solid black line). There is no massless mode in this set.}
\end{figure}

The criterium to select the relevant solutions, i.e. continuity and finiteness in the whole $y$ space,
indicates that for positive $\tilde{c}$
the mass spectra are in fact continua and not just discrete as obtained in the previous section.
In the quantum mechanics analog approach, the boundary conditions resulted
too restrictive to allow for the whole spectrum of the original problem and quantization
arose as a byproduct.
However, it does not mean that the actual spectrum belongs to the continuum
for every value of $\tilde{c}$.
After a numerical survey, we observe that for all $\tilde{c}\leq 0$  the mass spectra are discrete even
in the full approach (see e.g. Table \ref{table1} where we show all the first values of $m$
for $\tilde{c} = -1$; see also Fig. \ref{a_6_sym} and \ref{a_6_asym}).
For positive values, i.e. $\tilde{c}\in (0,6)$, 
on the other hand, arbitrary values of the mass allow nondivergent $w(y)$ solutions.
To illustrate this point, in Fig. \ref{a_6_1 sym} and \ref{a_6_1 asym}
we have chosen a generic value of $\tilde{c}$
to display the solutions for several arbitrary values of $m$.
For the special cases of $\tilde{c}$ studied in the previous Section
the results coincide, as expected. However, with the important difference
that on top of the previously found sequence the mass spectra now obtained
fulfill  it to the continuum, including zero.
%

%
%
%
\begin{center}
\begin{table}[hb]
\caption{List of the first values of $m_s$ (symmetric solutions) and $m_a$ (antisymmetric ones) for $\tilde{c}=-1$ ($\gamma=2$).}
\label{table1}
\vskip 0.2cm%
\begin{tabular}{ccc}
  \hline  \hline
\,   \, & $m_s$\, & \, $m_a$\\ \hline
 \,   \, & \, 0.0000000  \, & \, $--$ \\
  \,   \, & \,5.01256460  \, & \, 2.88880921 \\  
   \,   \, & \,9.11222614  \, & \, 7.07405270 \\  
    \,   \, & \,13.15839916 \, & \,11.13871772 \\  
     \,   \, & \,17.18603686 \, & \,15.17371323 \\  
      \,   \, & \,21.20478031 \, & \,19.19621044 \\  
       \,   \, & \,25.21848382 \, & \,23.21211936 \\  
        \,   \, & \,29.22902086 \, & \,27.22407315 \\  
         \,   \, & \,33.23742273 \, & \,31.23344577 \\  
          \,   \, & \,37.24430826 \, & \,35.24102973 \\  
           \,   \, & \,41.25007349 \, & \,39.24731247 \\  
            \hline
             \hline
\end{tabular}
\end{table}
\end{center}
\begin{figure}[ht]
\includegraphics[width=8cm,height=5.5cm]{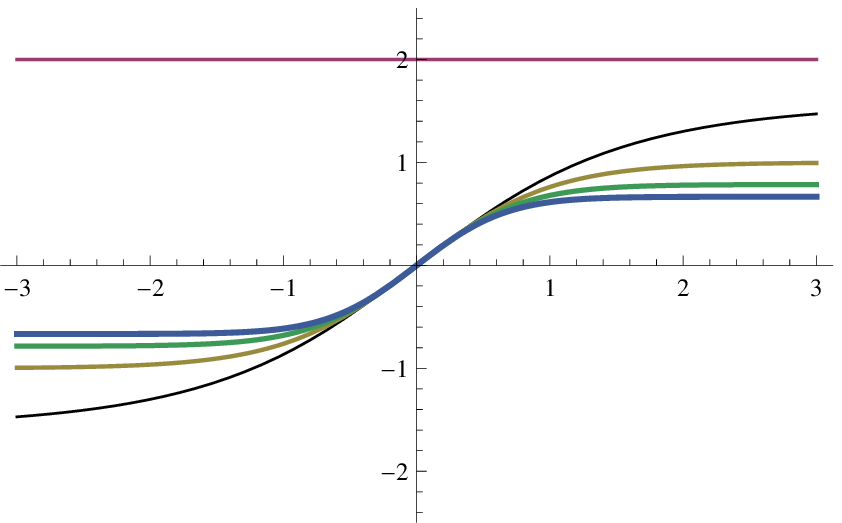}
\caption { \label{ctil_1to4} Some zero modes for $\tilde{c}>0$ (thinner to thicker):  $w_0^{\tilde{c}=1}(y)=\arcsin(\tanh \!y)$; $w_0^{\tilde{c}=2}(y)=\tanh y$; $w_0^{\tilde{c}=3}(y)=\frac{1}{2} \left[\sech\!y \tanh\!y+\arcsin(\tanh\!y) \right]$; $w_0^{\tilde{c}=4}(x)=\frac{1}{3} \tanh\!y (3-\tanh^2\!y)$. A constant zero mode is present for any $\tilde{c}$.}
\includegraphics[width=8cm,height=5.5cm]{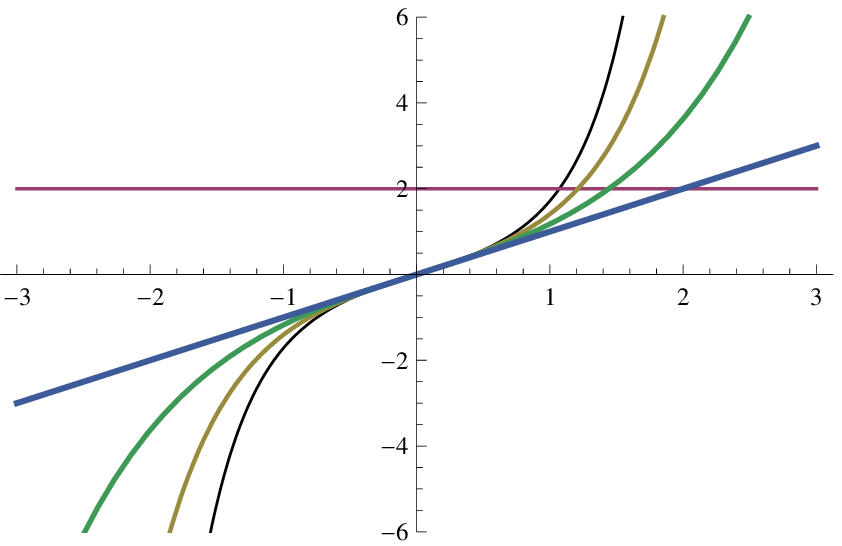}
\caption { \label{-ctil_0to3}  Some zero-modes for $\tilde{c}\leq0$ (thinner to thicker): $w_0^{\tilde{c}=-3}(y)=\frac{1}{3} \sinh\!y (3+\sinh^2\!y)$; $w_0^{\tilde{c}=-2}(y)=\frac{1}{2}(y+\sinh\!y \cosh\!y)$; $w_0^{\tilde{c}=-1}(y)=\sinh\!y$; $w_0^{\tilde{c}=0}(y)= y$. A constant zero-mode is present for any $\tilde{c}$.}
\end{figure}
\subsection{The zero-mode of the Kalb-Ramond field \label{sub zero}}
As a consequence of the previous analysis,
we conclude that the quantum analog Schrodinger -like problem returns just
a discrete cut of the spectrum and is therefore not fully appropriate to exhaustively solve the problem.
In some cases, the zero-mode is not even manifest 
 since eq. \ref{schrodinger}, with the appropriate boundary conditions at $z=\pm 1$,
does not allow such a solution (see Sect. \ref{subsub tilde2} and \ref{subsub tilde3}).
A direct approach, on the other hand, allows an analytical $m=0$ solution to
Eq. (\ref{massive modes 1}) for any value of $\tilde{c}$ :
\beq
w_0(y)=e_1 + e_2 \int^ y \sech^{\tilde{c}}\!y \,dy.  \label{eq2}
\eeq
After solving this integral the general zero-mode reads
\beq
w_0(y)=e_1 + e_2 \tanh\!y\ _2F_1\left(\frac{1}{2},\,
\frac{1-\tilde{c}}{2},\,\frac{3}{2}\,;\,\tanh^2\!y\right) \label{eq3}\eeq
%
which is finite everywhere provided $\tilde{c}$ is positive
(see Figs. \ref{ctil_1to4} and \ref{-ctil_0to3}). Gauss hypergeometric functions are
defined as
$_2F_1\left(a, b, c; x\right)= 1+\frac{a\ b\ x}{c\ 1 ! }+\frac{a(a+1)\ b(b+1) x^2}{c(c+1)\ 2 !}+\dots\ .$
The constant solution $w_0(y) = e_1$ represents the symmetric zero-mode (it is
of course finite and continuous for arbitrary $\tilde{c})$.

Thus, for $\tilde{c}>0$ there are
two independent zero-modes while for $\tilde{c}\leq0$ only the constant one.
This was the result expected in the Schrodinger approach, where the ground-state
must be symmetric. However, the associated boundary conditions
precluded the zero-mode and the allowed ground state emerged as a massive mode
(see e.g. Figs. \ref{a6_c2_par} and \ref{a6_c2_impar}).
Independently of the approach, the gain of mass of the ground state may suggest
unobservable the experimental signature of this field or, in turn, give important information
about the dilatonic coupling indicating
the exact value of the parameters of the model as we discuss in what follows.

\section{Final discussion \label{remarks}}

Although the completeness of the Schrodinger approach is
restricted to a particular range of parameters, it is plenty useful to evaluate the low energy
features of the system, especially for negative values of $\tilde{c}$.

At very low energies we can compute the corresponding
nonrelativistic effective potential to estimate the
Kaluza-Klein contributions to the effective coupling to matter.
Since the nonrelativistic potential, as given by
the Green function of the equation of motion integrated along the source, depends
on the square of the massive amplitudes at the origin, we must simply
compare their relative weights to assess their contributions, see \cite{jaume}.

We can assume that the coupling with the brane takes
place exactly on the 4D ordinary space-time, at $y=0$.
Since matter fields are constrained to the wall, it is
precisely there that the relevant physical effects should be more important.
As shown in Table \ref{table c=-1}, the normalized square amplitudes of
the KK modes are rapidly decreasing with mass.
In addition, each contribution to the effective nonrelativistic 4D static potential
presents an exponential factor of the Yukawa type which also depends on the mass of the mode,
further quenching the higher mass eigenstates (see also \cite{RS,csaki al} for a study of
this issue in the case of the gravitational field).
\begin{center}
\begin{table}[hb]
\caption{List of first mass eigenstate square amplitudes at the origin;
  $\tilde{c}=-1$.}
\label{table c=-1}
\vskip 0.2cm%
\begin{tabular}{ccc}
  \hline  \hline
    & $m_s$\, & \, $w_m^2(0)$\\ \hline
     & 0           \, & \, 1.00000000000   \\
      & 5.01256460  \, & \, 0.35558755410   \\  
       & 9.11222614  \, & \, 0.19248779813  \\   
        & 13.15839916 \, & \, 0.13482515859   \\  
         & 17.18603686 \, & \, 0.10488449142   \\  
          & 21.20478031 \, & \, 0.08640457375   \\  
           & 25.21848382 \, & \, 0.07379433274   \\  
            & 29.22902086 \, & \, 0.06461526991   \\  
             & 33.23742273 \, & \, 0.05761117479   \\ 
              & 37.24430826 \, & \, 0.05208093848   \\  
               & 41.25007349 \, & \, 0.04759600275   \\  
            \hline
             \hline
\end{tabular}
\end{table}
\end{center}
Thus, although KK excitations have nonvanishing momenta in
the fifth direction interactions mediated by massive modes
will be strongly suppressed near the brane\-world.

Since the axion mass is expected to be very light but nonzero
($10^{-6}<m_\chi[\rm eV]<0.01$) \cite{nature06}, this particle would be just faintly
coupled to matter fields  and therefore hardly detectable by direct means \cite{matter axion}.
Although the experimental issues, several important collaborations are currently
searching for an axion signal \cite{exp search axion}.

{In the present scenario, assuming
that the axion is indeed a physical realization of the KR field,
the eventual advent of an experimental result for the axion mass
would suggest a value for the $\tilde{c}$ parameter in our model.
A way to see this could be in principle accomplished by means of
a detailed (and lengthy) inspection of a sufficiently large
number of spectra for different values of $\tilde{c}$
until one finds the eigenvalue that better fits the experimental mass.
For this, it would be necessary to fix the $b$ parameter in first place.
If the axion happened to be massless we could associate a positive $\tilde{c}$.
In this case, however, this prediction would be not very useful since
we would not know exactly which particular value of $\tilde{c}$ between 0 and 6
(remember that for every positive $\tilde{c}$ there is a zero mode and no gap).
As a matter of fact, this possibility is unlikely in face of the
current experimental expectancy of a light but nonzero mass axion. Indeed,
assuming that a mass gap exists in the physical spectrum of this particle,
our model would associate a negative $\tilde{c}$ to the experimental mass, as explained above.
For each value of $\tilde{c}$ we can compare its minimum eigenvalue to the
axion mass. If we first transform $y\rightarrow \kappa y$ in eq.\ref{massive modes},
which results in a scaling of the mass $m_n^2\rightarrow m_n^2/\kappa^2\equiv m_\chi^2$, we will get
a range of values of $ \kappa $ for a range in $m_\chi$.
For instance, let us choose $\tilde{c}=-1$
so that $m_1= 5.01256460$. For an axion mass in $10^{-6}< m_\chi[\rm eV]< 10^{-2}$ then
$9.9174 \times 10  \approx 100 >\kappa^{-1}[\rm cm]>9.9174 \times 10^{-5} \approx 0.0001$,
characterizing the size of the scale transformation.
Now, since this parameter is defined as $\tilde{c} =-3(15+4\lambda\sqrt{3M^3})$
one could thereon get a theoretical prediction of the product $\lambda M^{3/2}$.
Recall that the 5D Planck mass $M$ depends $M_P$ and $b$ through eq.\ref{Mb}.}
%
%

%
%

The study of the interaction of this field with the photon \cite{axion-photon}
by means of a mixed Maxwell-Kalb-Ramond lagrangian
would allow for an additional theoretical tool to bear the axion problem.
If it is true that the Kalb-Ramond axion supplies the spacetime with torsion in its nearness,
a kind of primordial optical activity would be possible.
Indeed, a non-Faraday rotation, independent of the wavelength and parameters of the source
and intergalactic medium, has already been predicted in \cite{indianos EPJ 2002}.
An optical rotation of the plane of polarization
of light coming from distant sources \cite{rotation} might then be observed in some near future.
A signal of this effect would provide evidence of the existence of a primordial KR field 
and hence of a likely component of cold dark matter in the universe.
This is being subject of further investigation.

\end{document}